\documentclass[onecolumn, preprintnumbers, showpacs, nofootinbib, aps]{revtex4-1}
\usepackage{graphicx}

\newcommand{\bi}{\bibitem}
\newcommand{\be}{\begin{eqnarray}}
\newcommand{\ee}{\end{eqnarray}}

\begin{document}

\title{3D simulations of the accretion process in Kerr space-time 
with arbitrary value of the spin parameter}

\author{Cosimo Bambi}
\email{cosimo.bambi@ipmu.jp}

\author{Naoki Yoshida}
\email{naoki.yoshida@ipmu.jp}

\affiliation{
Institute for the Physics and Mathematics of the Universe, 
The University of Tokyo, Kashiwa, Chiba 277-8583, Japan}

\date{\today}

\preprint{IPMU10-0097}

\begin{abstract}
We present the results of three-dimensional
general relativistic hydrodynamic simulations of adiabatic and
spherically symmetric accretion in Kerr space-time. We consider 
compact objects with spin parameter $|a_*| \le 1$ (black 
holes) and with $|a_*| > 1$ (super-spinars). Our full three-dimensional
simulations confirm the formation of equatorial outflows for high values 
of $|a_*|$, as found in our previous work in 2.5 dimensions.
We show that the critical value of $|a_*|$ determining the onset of powerful outflows 
depends mainly on 
the radius of the compact object. The phenomenon of equatorial 
outflows can hardly occur around a black hole and may thus 
be used to test the bound $|a_*| \le 1$ for astrophysical
black hole candidates.
\end{abstract}


\maketitle


\section{Introduction}

It is widely believed that the final product of gravitational
collapse is a Kerr black hole: an object with an event horizon 
that is completely characterized by two parameters, the mass $M$ and 
the spin $J$. The condition for the existence of the event horizon
is $|a_*| \le 1$, where $a_* = J/M^2$ is the dimensionless spin 
parameter. The black hole paradigm relies on three ingredients. 
The first is that, in general relativity, under apparently 
reasonable assumptions, the collapsing matter forms space-time
singularities~\cite{hawking}. The second one is the Cosmic Censorship
Conjecture~\cite{penrose}, according to which all the space-time
singularities must be hidden behind an event horizon. The last 
ingredient is the fact that in 4-dimensional general relativity 
the Kerr space-time is the only asymptotically-flat and stationary 
solution with a regular event horizon~\cite{carter, robinson}.

The Cosmic Censorship Conjecture is a simple requirement to get 
rid of pathological space-times with unphysical properties. 
However, it seems to be motivated by our poor knowledge of 
the theory at high energy rather than by true physical reasons. 
Space-times with naked singularities may indeed arise from the 
break down of our theories at high energy. Instead of naked 
singularity, we might thus talk about effective naked 
singularity~\cite{nakao, harada}: a space-time region outside 
an event horizon in which classical general relativity cannot be 
used and the future predictability is lost without the theory 
of quantum gravity. It is also remarkable that we know several
counter-examples violating the Cosmic Censorship Conjecture (see 
e.g. Refs.~\cite{chr84, piran, joshi93, chr94, joshi94, joshi98}).

In absence of an event horizon, there is no uniqueness theorem
and, in general, the final product of the collapse is not as 
simple as a Kerr black hole. Nevertheless, as a first approximation,
the space-time around the compact object can still be described 
by $M$ and $J$ if higher order multipole moments are subdominant. 
{\it The simplest test to rule out the Kerr black hole as the 
final product of the gravitational collapse is thus to find a 
massive and compact object with $|a_*| > 1$ (super-spinar)}. The 
idea that the end state of the collapse could violate the bound
$|a_*| \le 1$ was suggested in~\citep{horava}. For the possibility
of over-spinning an existent black hole, see e.g. Ref.~\cite{ted}. 
The properties of the electromagnetic radiation emitted around 
Kerr super-spinars were studied in~\cite{bf09, bft09, th10} and 
compared with the case of Kerr black hole, in order to examine 
how to test the bound $|a_*| \le 1$ with future experiments.

The accretion process onto Kerr super-spinars in 2.5 dimensions
(2 spatial dimensions, vectors with 3 spatial components) has been 
studied in~\cite{bfhty09, b09, bhty10}. Close to the object, 
the gravitational force can be repulsive. In Ref.~\cite{bhty10}, 
we discussed the Bondi accretion and found the formation of 
equatorial outflows. These outflows are very different from the 
one expected in the accretion process onto black holes: they are 
produced by the repulsive gravitational force at short distances, 
rather than by magnetic fields, and the gas is ejected on and 
around the equatorial plane, while in the black hole case the outflows are 
expected parallel to the spin. Current observations cannot
directly probe if outflows and jets observed in black hole candidates are
perpendicular or parallel to the spin of the massive objects,
most of jets from radio galaxies appear to be of dipolar 
shape, being very different from planar outflows.

In this paper, we extend the work of Ref.~\cite{bhty10}. We study the 
accretion process in 3 spatial dimensions. The main aim is to
understand the structure of these outflows on the equatorial 
plane, which cannot be investigated in 2.5 dimensions. Indeed, 
one may naively expect roughly the following three basic configurations:
${\it i)}$ a perfect axial symmetry, in which the gas is ejected
isotropically in all the directions, ${\it ii)}$ the formation 
of a certain number of collimated and stable outflows on the
equatorial plane, or
${\it iii)}$ a strongly chaotic phenomenon, in which the accreting 
material is ejected randomly along several directions and there is 
no formation of stable outflows. Our simulations suggest the
third case.

The paper is organized as follows. In Sec.~\ref{s-model}, we 
discuss the model for super-spinars. In Sec.~\ref{s-results}, 
we present the results of our 3-dimensional general relativistic
hydrodynamic (3D GRHD) simulations. In particular, we show the 
cases of Schwarzschild black hole ($a_* = 0$), extreme Kerr black 
hole ($a_* = 1$), and Kerr super-spinar with $a_* = 2$ and 3. In 
Sec.~\ref{s-disc}, we discuss the results. In Sec.~\ref{s-conc}, 
there are summary and conclusions. Throughout the paper we use 
Boyer-Lindquist coordinates to describe the Kerr background and 
natural units $G_N = c = k_B = 1$.

\section{Super-spinar model \label{s-model}}

The Kerr space-time with $|a_*| > 1$ has a naked singularity at
$r = 0$. For $r > 0$, the space-time is everywhere regular. The
singularity at $r = 0$ is point-like, but has the topology of a 
ring: it can connect our Universe, where $r > 0$, with another 
Universe, at $r < 0$. The problem is that an observer at $r > 0$ 
can go to the other Universe and, in absence of an event horizon, 
come back to our Universe at an earlier time; that is, the naked
singularity can be used as a time machine~\cite{carter2}. However, 
one may expect that general relativity breaks down before reaching 
the singularity and that eventually causality cannot be violated. 
In absence of the complete theory of quantum gravity, we have to take a
phenomenological approach to study the astrophysical properties of 
super-spinars. Here we consider Kerr super-spinars and we assume 
they can be modeled as a massive object with a radius 
$r \approx 2.5 \, M$, whose surface absorbs all the accreting 
matter hitting it (the possibility
of a surface with different properties is briefly discussed in
Sec.~\ref{s-disc}). Our model is motivated by the following
simple considerations.
\begin{enumerate}
\item Roughly speaking, pathological regions with closed time-like 
curves can be removed by exciting the space-time. This occurs, for 
example, if the space-time goes to a new phase and a domain wall is 
formed~\cite{horava2, israel, drukker, gimon}. Across the domain 
wall, the metric is non-differentiable and the expected region with 
closed time-like curves arises from the naive continuation of the 
metric ignoring the domain wall. The latter can be made of very 
exotic stuff, e.g. supertubes~\cite{horava2, drukker, gimon} or 
fundamental strings~\cite{israel}.
\item A crucial point is the radius of the new object. Here we
take $r \approx 2.5 \, M$ in order to prevent the instability of
the super-spinar~\cite{enrico} and have all the thermodynamical
variables under control. For smaller/larger radii, one can properly
rescale (decrease/increase) the spin parameter to obtain a very
similar accretion process. For astrophysical bodies, $M$
is a length much larger than the Planck scale $L_{Pl} \sim 10^{-33}$~cm,
where quantum gravity effects are often expected. However, the
gravitational radius $r_g \sim M$, rather than $L_{Pl}$, 
may be the critical quantity. This occurs, for example, in
the fuzzball picture~\cite{mathur}: 
here black holes have no horizon and the 
central singularity is replaced by a long string spreading over 
the volume classically occupied by the black hole. Even if so far 
only a few very special black holes have been studied, if correct, 
the fuzzball picture should be applied (somehow) even to the 
final product of the collapse of a star. The latter may thus 
resemble more a star made of very exotic matter than a black hole. 
However, unlike recent proposals such as boson stars, general 
relativity may not hold inside the object.
\item In this work, we use the Kerr metric, which is described only 
by two parameters, $M$ and $J$. In absence of an event horizon, one
should instead expect a more complex object. Even if that is true,
$M$ and $J$ are more likely the dominant terms in the multipole
moment expansion and determine the main features of the accretion 
process. 
\end{enumerate}

\section{3D simulations \label{s-results}}

In this section, we present the results of our 3D simulations in
Kerr space-time with arbitrary value of the spin parameter $a_*$;
for previous works on the accretion process in full general
relativity, see e.g. Refs.~\cite{hsw1, hsw2, piran2, chak}.
The code is a revised version of the one used 
in~\cite{bfhty09, b09, bhty10}. Now the (default) computational 
domain is $2.5 \, M < r < 40 \, M$, $0 < \theta < \pi$, and 
$0 < \phi < 2\pi$. Unlike in~\cite{bfhty09, b09, bhty10}, 
the thermodynamical variables are everywhere under control and 
we do not need to impose a maximum temperature. The accretion 
process is adiabatic. We assume an ideal non-relativistic gas
with $\Gamma = 5/3$, which is injected into the computational 
domain from the outer boundary at a constant rate and isotropically.
We also examine several cases with $\Gamma = 4/3$ for relativistic
particles. 
The gravitational field of the accreting matter is neglected (test 
fluid approximation). The initial conditions are the same of
Refs.~\cite{bfhty09, b09, bhty10}. 
In the next subsections, we present the results 
of the simulations for a few different values of the spin parameter.
The simulations run from $t = 0$ to $t = 500 \, M$ (for $a_* = 0$, 1,
and 2) or $t = 300 \, M$ (for $a_* = 3$). 
We did not find a quasi-steady-state configuration
within this period.
However, it seems to be enough to catch the main features of the accretion 
process.

\subsection{Schwarzschild black hole ($a_* = 0$)}

\begin{figure}
\par
\begin{center}
\includegraphics[height=7cm,angle=0]{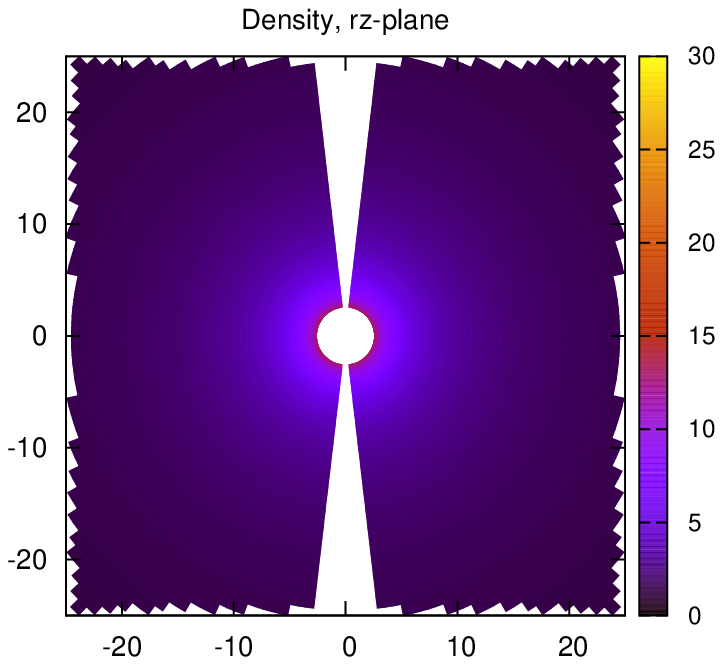} \hspace{.3cm}
\includegraphics[height=7cm,angle=0]{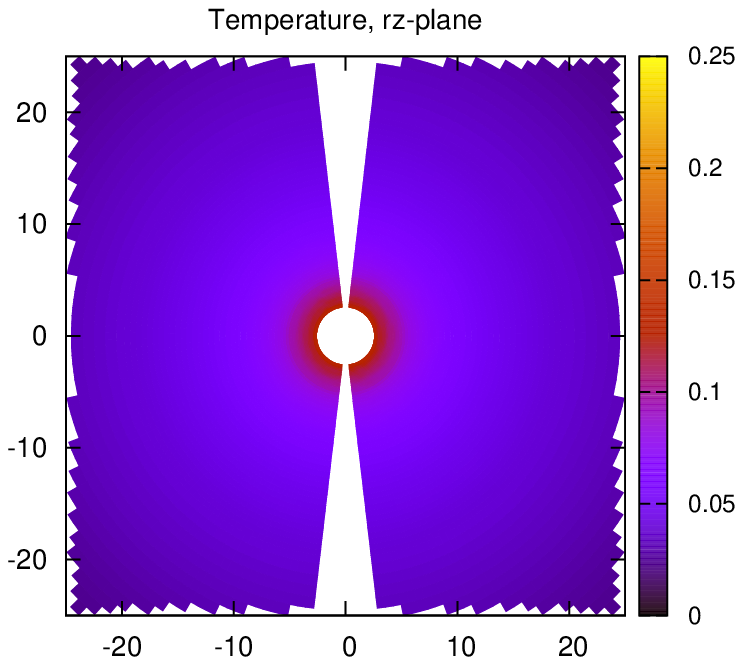} \\
\includegraphics[height=7cm,angle=0]{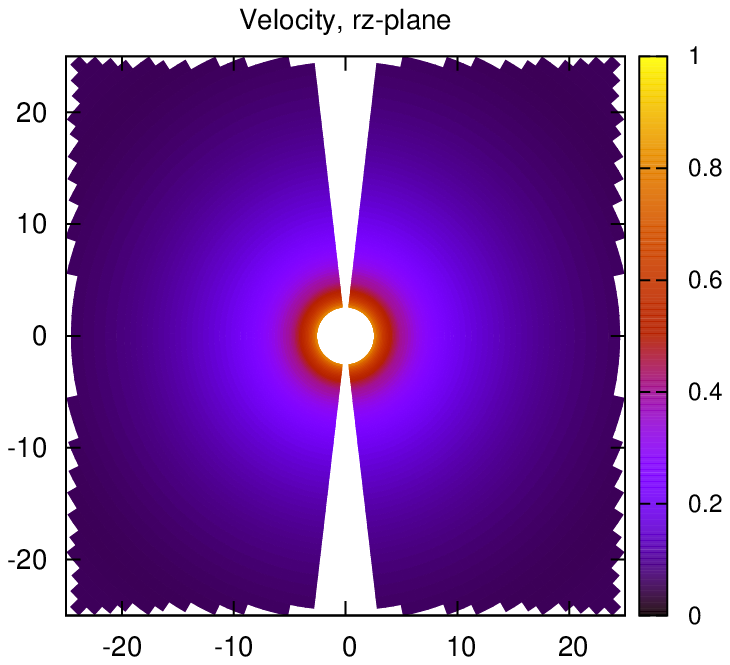} \\
\includegraphics[height=7cm,angle=0]{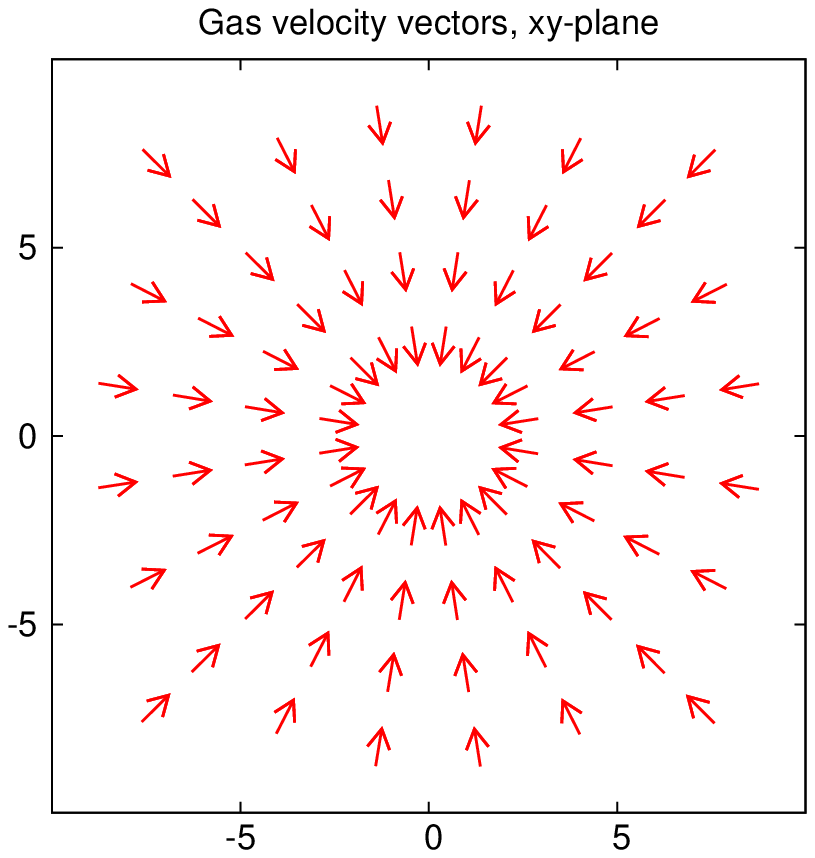}
\end{center}
\par
\vspace{-5mm} 
\caption{We plot the density $\rho$ (in arbitrary units),  
the temperature $T$ (in GeV), and the velocity 
$v = \sqrt{\gamma_{ij}v^iv^j}$ of the accreting 
gas around a Schwarzschild black hole ($a_* = 0$) 
at $t = 500 \, M$. The bottom 
panel shows the direction of the gas velocity. The $xy-$plane 
is the equatorial plane, while the $rz-$plane is a plane 
containing the axis of symmetry $z$. The unit of length along 
the axes is $M$.}
\label{f-spin0}
\end{figure}

This is the simplest case. The accretion process is almost perfectly
spherically symmetric, see Fig.~\ref{f-spin0}. At the inner boundary 
$r_{in} = 2.5 \, M$ and at $t = 500 \, M$, our code predicts a 
temperature $T \approx 125$~MeV and a velocity 
$v = \sqrt{\gamma_{ij} v^i v^j} \approx 0.8$, where $\gamma_{ij}$ 
is the 3-metric.

\begin{figure}
\par
\begin{center}
\includegraphics[height=5.7cm,angle=0]{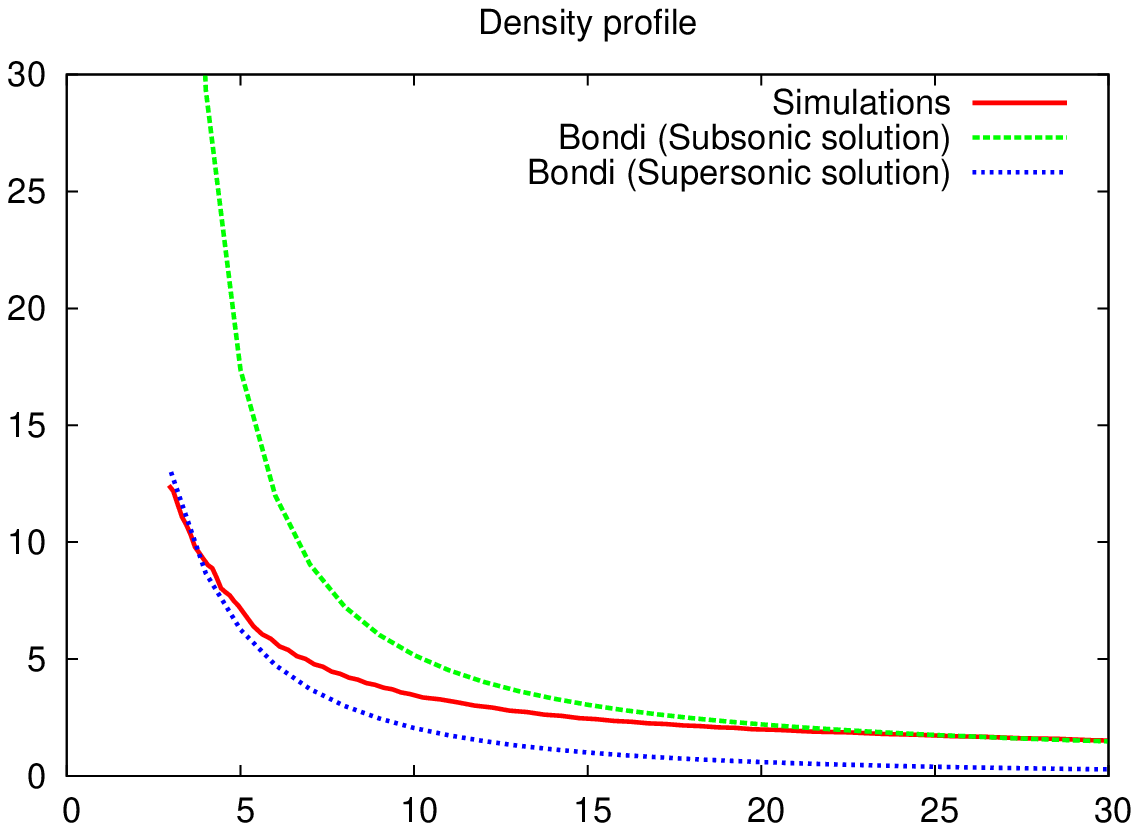} \hspace{.3cm}
\includegraphics[height=5.7cm,angle=0]{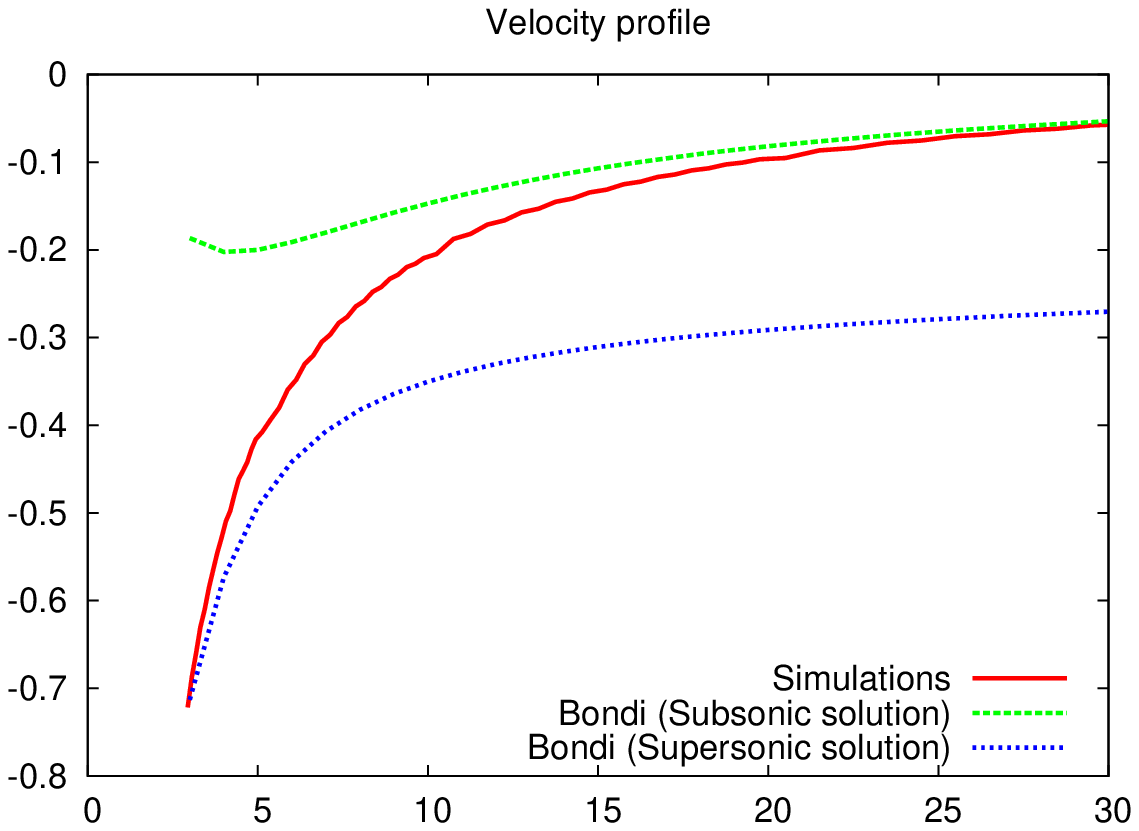} \\
\includegraphics[height=5.7cm,angle=0]{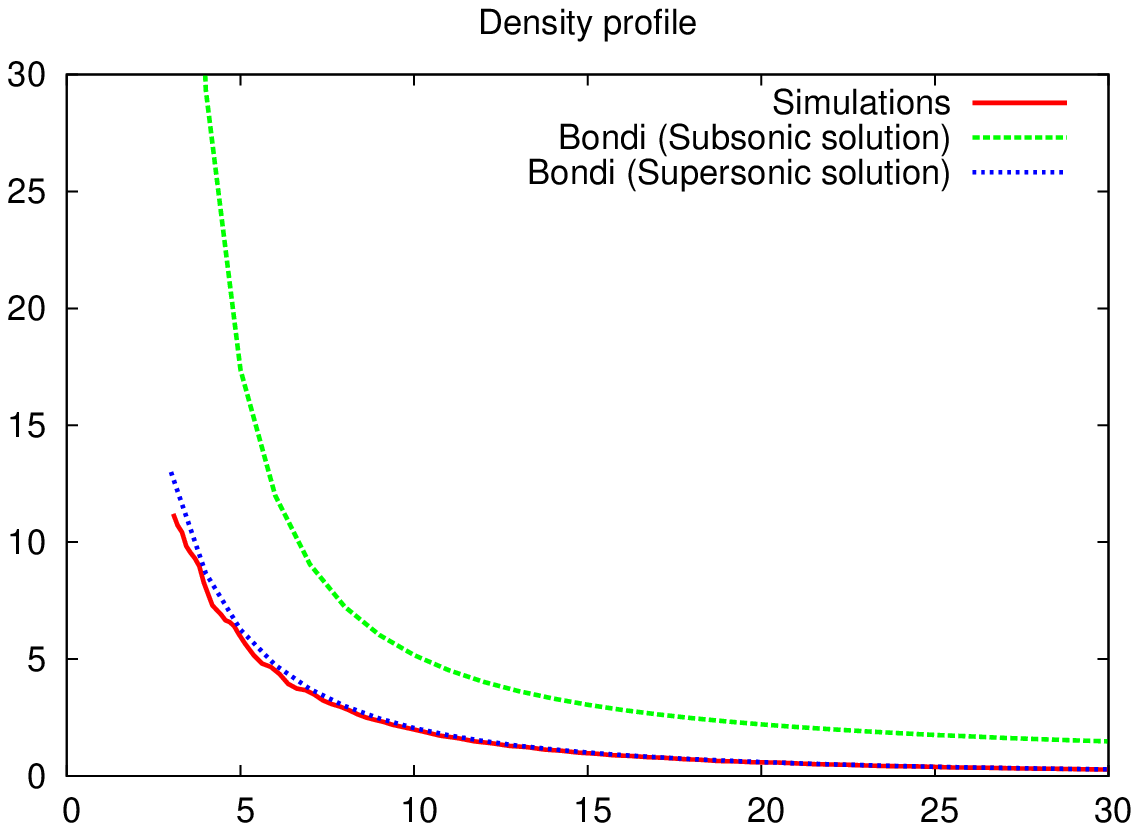} \hspace{.3cm}
\includegraphics[height=5.7cm,angle=0]{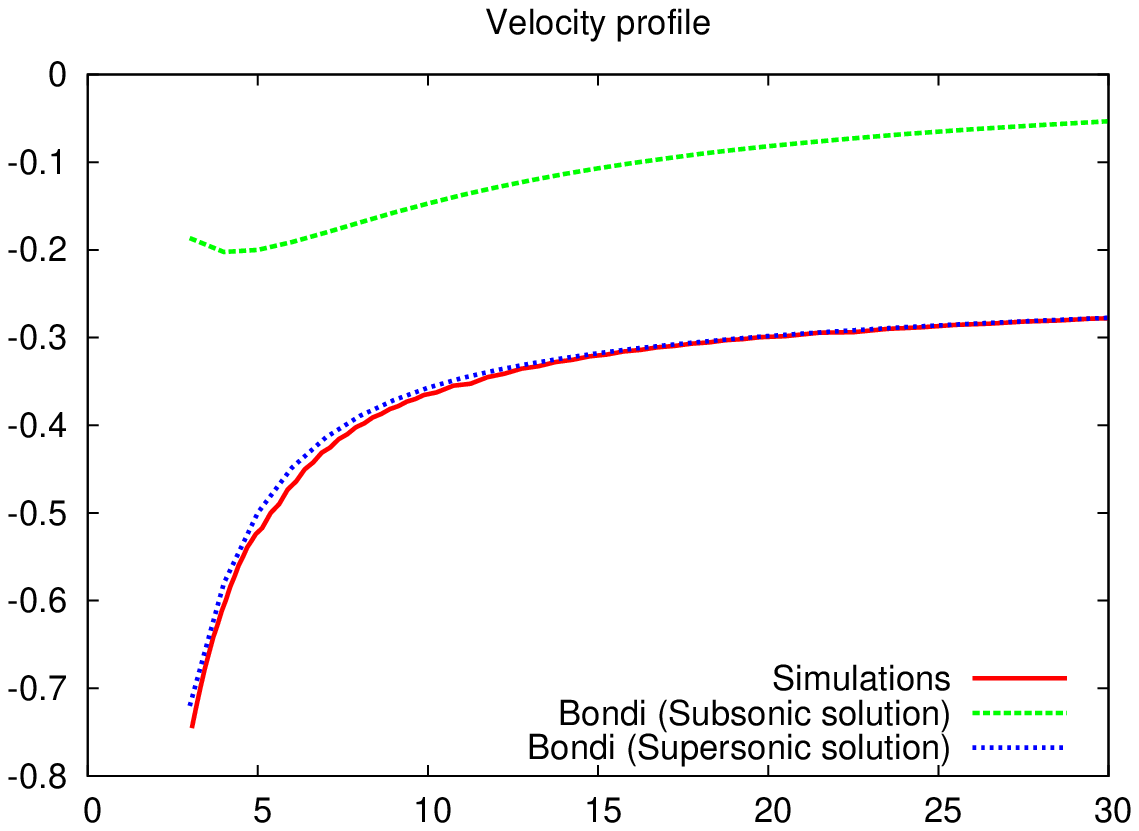}
\end{center}
\par
\vspace{-5mm} 
\caption{Comparison of the density profile and of the radial
velocity profile (as seen by a local observer) for $a_* = 0$, 
between our numerical simulations at $t = 500 \, M$ and the
semi-analytical solutions of the relativistic Bondi equations.
Assuming the initial gas distribution adopted in the simulations 
of this work, we find that the numerical solution is close to
the supersonic solution at small radii and to the subsonic
solution at larger radii (top panels). If we take the computational
domain initially empty and we inject gas from the outer
boundary, we can find a numerical solution close to the
supersonic solution everywhere (bottom panels).}
\label{f-cfr-bondi}
\end{figure}

For a Schwarzschild black hole, the general relativistic Bondi solutions 
are well known~\cite{michel} (see also Appendix G of Ref.~\cite{st-book}).
The master equations are
\be
4 \pi r^2 \rho u &=& constant \, , \\
\left(1 - \frac{2M}{r} + u^2\right) h^2 &=& constant \, , 
\ee
together with an equation of state. 
Here $\rho$ is the rest-mass energy density,
$h = 1 + \epsilon + p/\rho$ the specific enthalpy, $\epsilon$
the specific internal energy density, 
$p$ the pressure, and $u = - u^r$ with
$u^r$ the radial component of the 4-velocity of the gas. 
We can thus compare the results of our numerical simulations with 
the Bondi solutions. Note, however, that the two results do not have to match
perfectly everywhere, because they are set up in a different way: 
in our simulations, we assume an initial gas distribution, 
we evolve the system, and we find a quasi-steady-state flow (for $a_* = 0$, 
this is achieved at $t \approx 300 \, M$); in the Bondi case, one 
assumes from the beginning that the flow is in a quasi-steady-state
configuration. The comparison is shown in Fig.~\ref{f-cfr-bondi}.
The initial configuration assumed in this work is a static cloud
of gas around the massive object. We then inject gas from the
outer boundary at a constant rate. Interestingly, the quasi-steady-state solution
we found is close to the supersonic Bondi solution near the black
hole, and approaches the subsonic Bondi solution near the outer
boundary (Fig.~\ref{f-cfr-bondi}, top panels). A numerical solution
quite similar to the Bondi ones can instead be reached by assuming that
the computational domain is initially empty, and by injecting gas 
from the outer boundary at a proper density, pressure, and velocity 
(Fig.~\ref{f-cfr-bondi}, bottom panels). The close agreement 
between the simulation results and the analytic solution
is rather encouraging.

\subsection{Extreme Kerr black hole ($a_* = 1$)}

\begin{figure}
\par
\begin{center}
\includegraphics[height=7cm,angle=0]{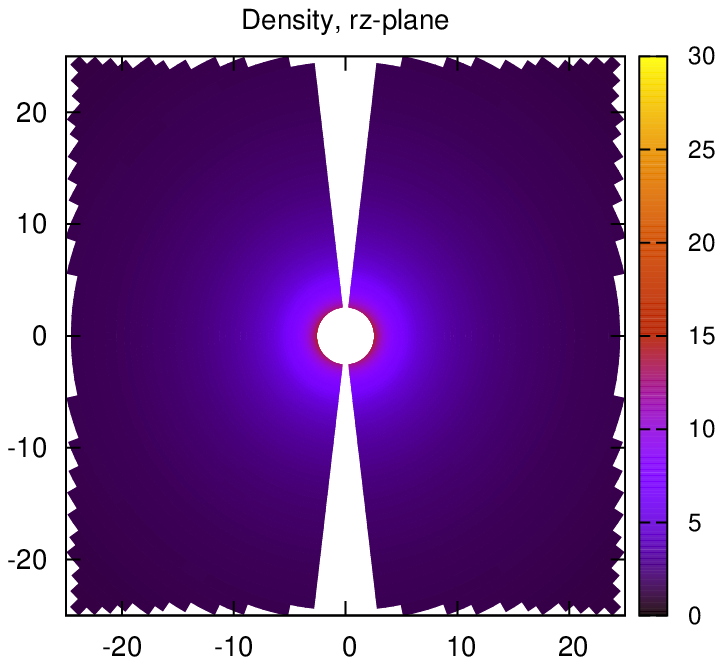} \hspace{.3cm}
\includegraphics[height=7cm,angle=0]{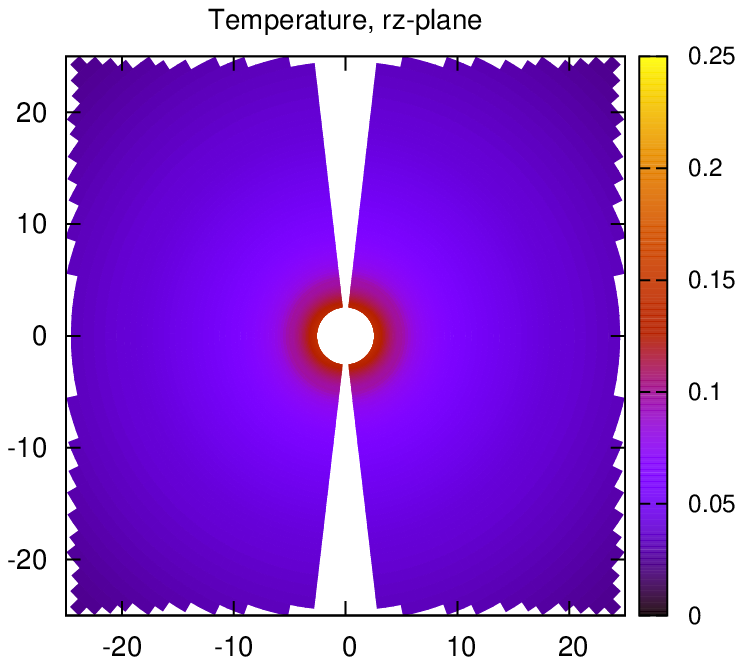} \\
\includegraphics[height=7cm,angle=0]{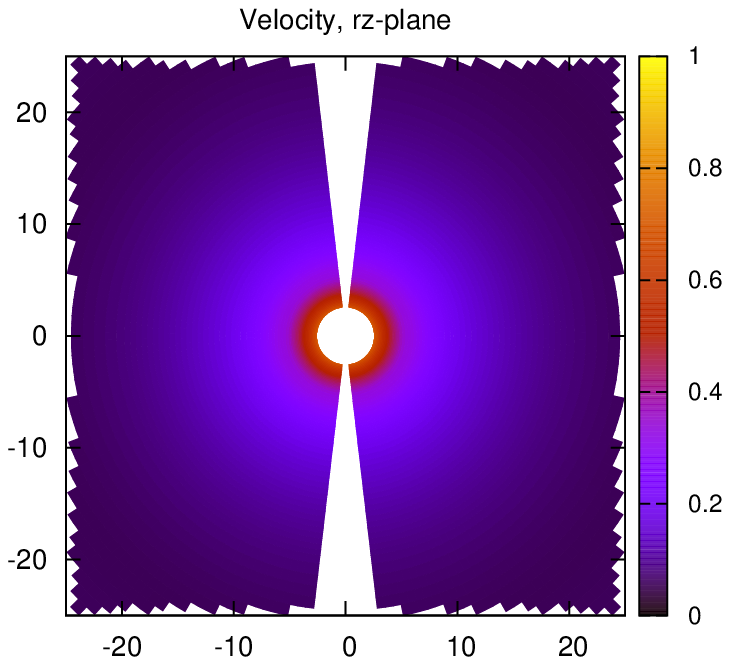} \\
\includegraphics[height=7cm,angle=0]{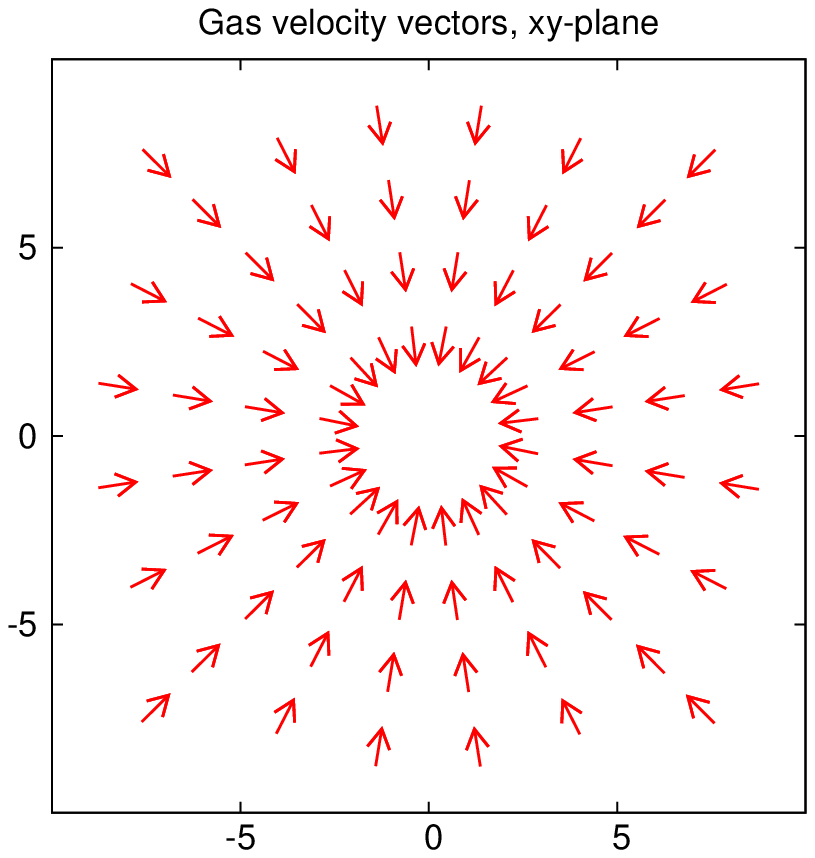}
\end{center}
\par
\vspace{-5mm} 
\caption{As in Fig.~\ref{f-spin0}, but for an extreme 
Kerr black hole ($a_* = 1$).}
\label{f-spin1}
\end{figure}

As we can see from Fig.~\ref{f-spin1}, the accretion process
onto an extreme Kerr black hole is still very similar to the 
Schwarzschild case. Now the density, the temperature, and the 
velocity of the gas are a little higher near the equatorial
plane than near the axis of symmetry; however, the 
difference can be hardly appreciated. At $r_{in}$, the 
temperature is in the range $130 - 140$~MeV and the velocity 
is $v \approx 0.7$. The latter is lower than the previous 
case, but we have also to consider that now the horizon is at 
$r_H = M$, while, for $a_* = 0$, we had $r_H = 2 \, M$. 
Let us notice that the computational domain does not include 
the ergoregion and that the velocity of the gas is essentially radial 
even near the inner boundary. The volume between $r_{in}$ and the 
horizon $r_H$ is, however, very small, and thus would not introduce any new 
interesting observational feature; indeed, most of the works 
in the literature use a pseudo-Newtonian potential, which 
neglects the spin of the massive object, and excludes the ergoregion
from the computational domain. It is important to notice
that in the case of black hole, the
thermodynamical variables at any radial coordinate $r$ are
independent of the choice of $r_{in}$. To be more precise,
that is generically true when the flow around the massive object
is supersonic; that is, $r_{in} < r_{s}$, where $r_s$ is the radius
of the sonic point. Here, at $t = 500 \, M$, we find 
$r_s \approx 6 \, M$.

\subsection{Kerr super-spinar with $a_* = 2$}

\begin{figure}
\par
\begin{center}
\includegraphics[height=7cm,angle=0]{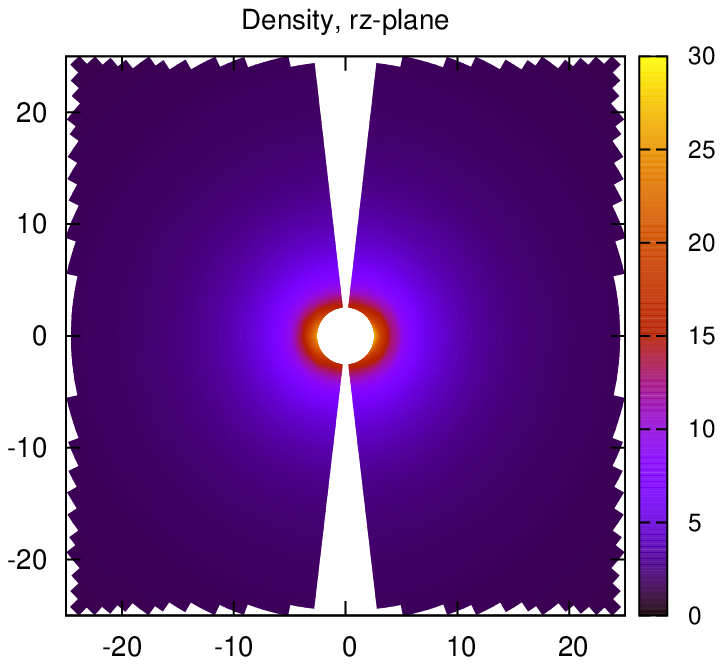} \hspace{.3cm}
\includegraphics[height=7cm,angle=0]{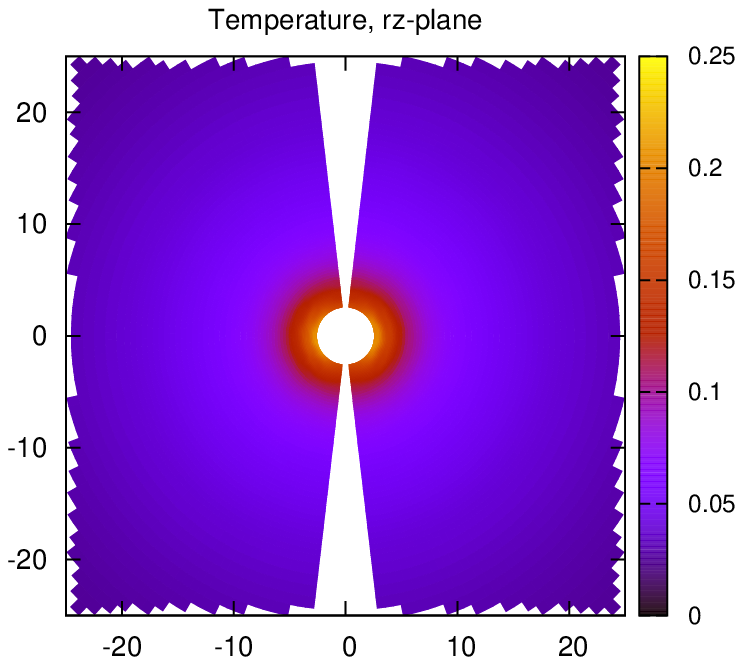} \\
\includegraphics[height=7cm,angle=0]{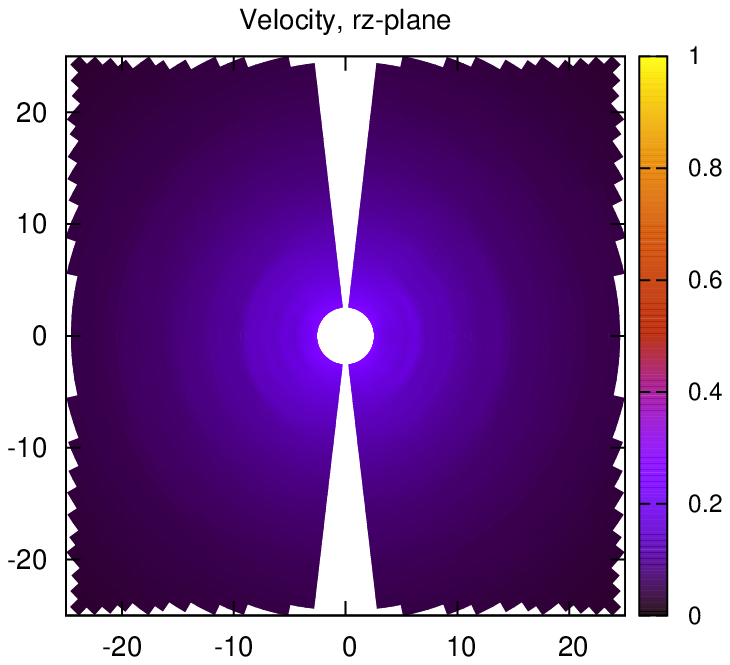} \hspace{.3cm}
\includegraphics[height=7cm,angle=0]{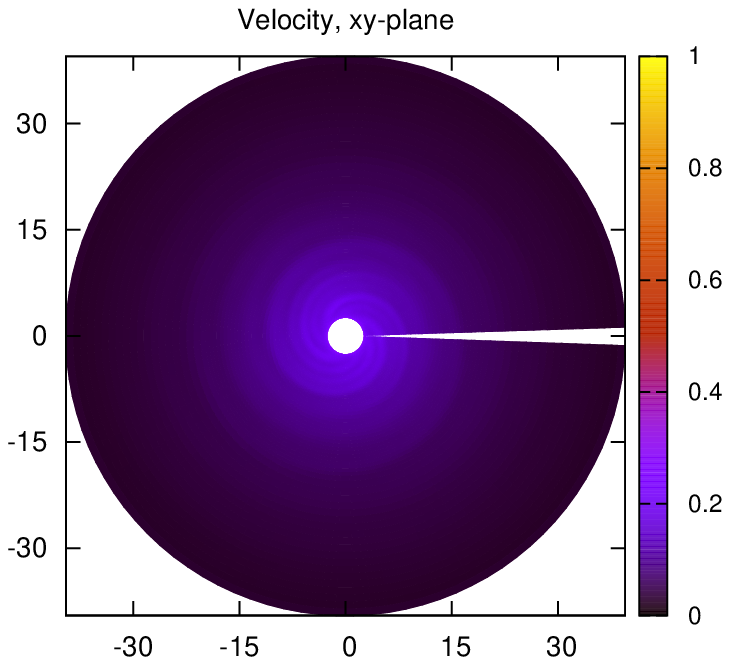} \\
\includegraphics[height=7cm,angle=0]{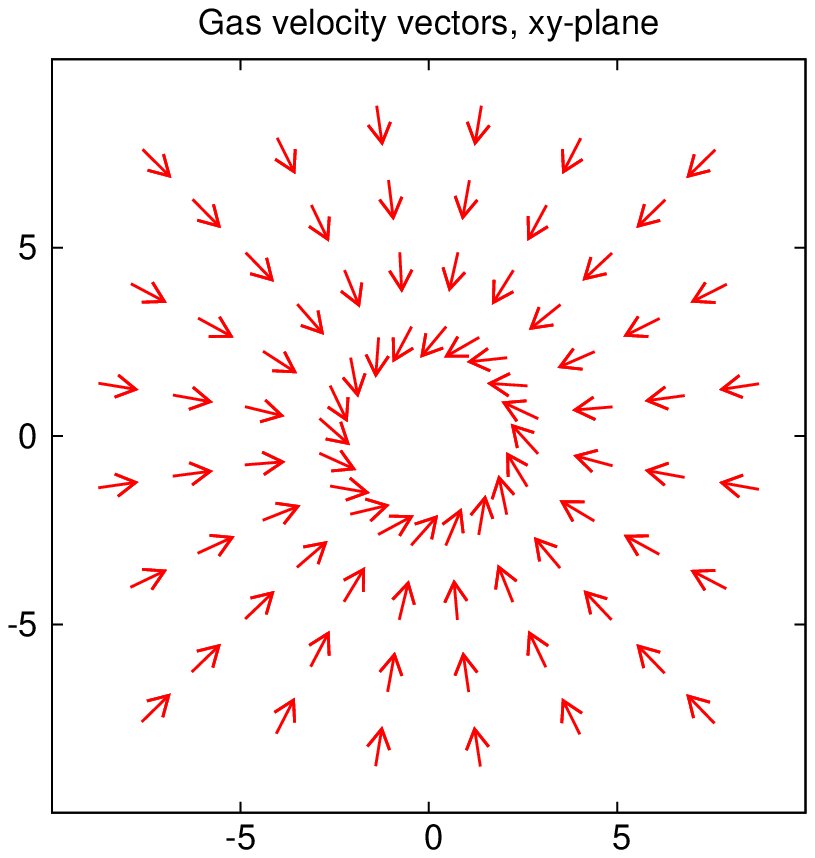}
\end{center}
\par
\vspace{-5mm} 
\caption{As in Fig.~\ref{f-spin0}, in the case of a Kerr 
super-spinar with $a_* = 2$.}
\label{f-spin2}
\end{figure}

\begin{figure}
\par
\begin{center}
\includegraphics[height=7cm,angle=0]{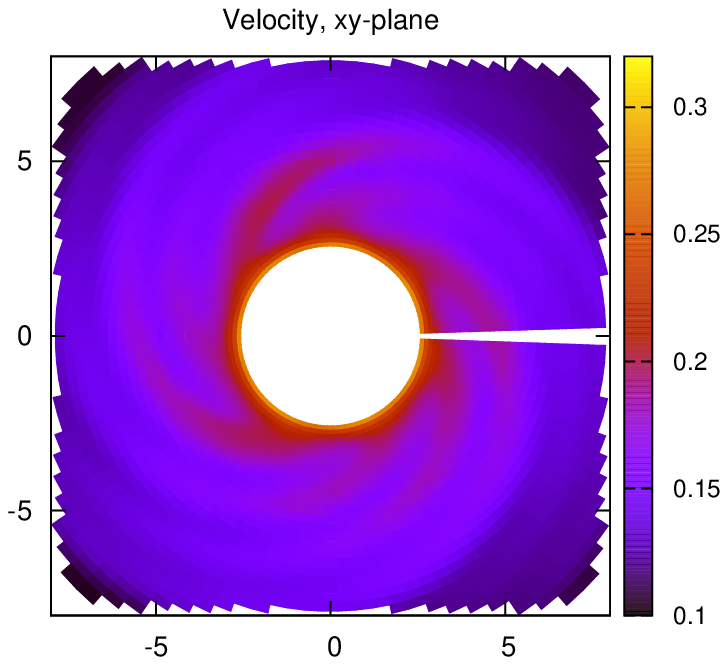}
\end{center}
\par
\vspace{-5mm} 
\caption{For $a_* = 2$, we find stable inflows forming a spiral
structure. Such a feature is essentially absent when we plot the
density or the temperature of the gas.}
\label{f-zoom}
\end{figure}

The accretion process for $a_* = 2$ is shown in Fig.~\ref{f-spin2}.
For $r_{in} = 2.5 \, M$, we exclude from the computational domain
the region with repulsive gravitational force\footnote{As discussed
in~\cite{bfhty09, b09}, the region around the super-spinar with
repulsive gravitational force is roughly given by the expression
$r < M |a_*| |\cos\theta|$.}. Despite that, the force is
weaker than around a black hole and the accretion process is 
inefficient. The result is that the velocity of the accreting
gas is lower, while the density and the temperature are higher.
At the inner boundary $r_{in}$, we find that the velocity $v$ is 
about 0.3 and the temperature is in the range $160 - 200$~MeV.
Now small variations of $r_{in}$ can change the accretion
process, because the flow is subsonic and the
thermodynamical variables at larger radii can depend on the
ones at smaller radii. As shown in the bottom panel of 
Fig.~\ref{f-spin2}, near the center the effect of the spin is 
relevant and the gas is forced to corotate with the super-spinar. 
In particular, we observe stable inflows forming a
spiral structure, see Fig.~\ref{f-zoom}.

\subsection{Kerr super-spinar with $a_* = 3$}

\begin{figure}
\par
\begin{center}
\includegraphics[height=7cm,angle=0]{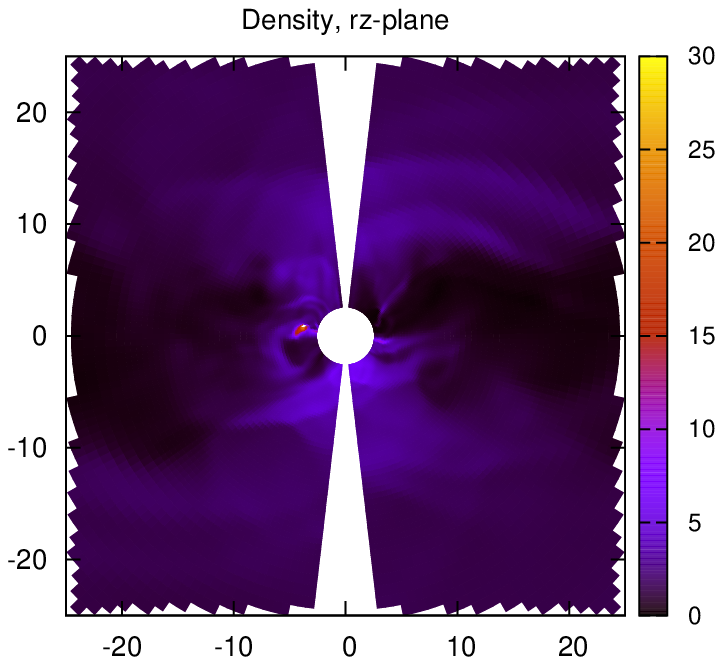} \hspace{.3cm}
\includegraphics[height=7cm,angle=0]{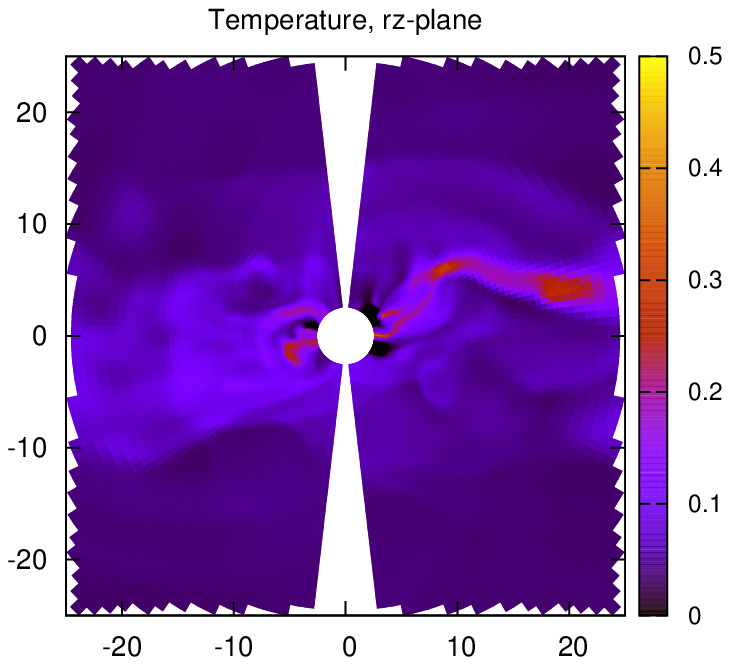} \\
\includegraphics[height=7cm,angle=0]{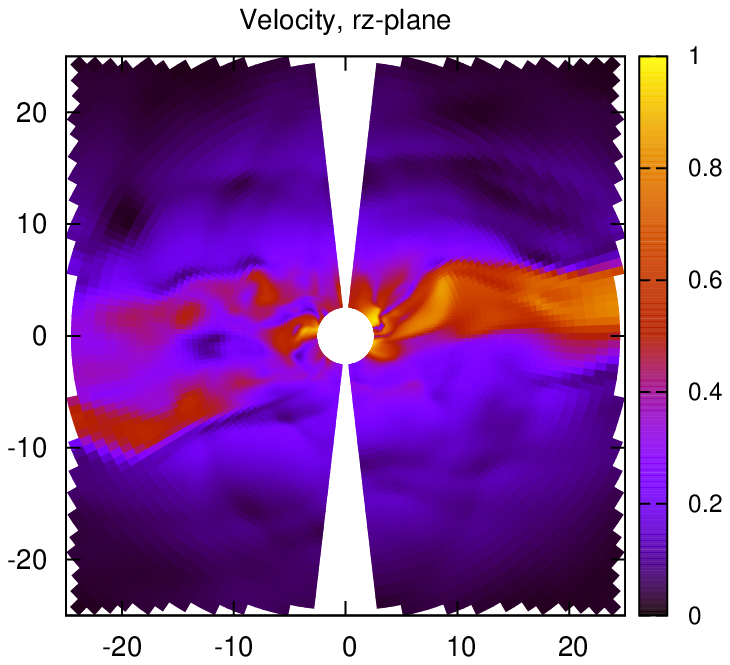} \hspace{.3cm}
\includegraphics[height=7cm,angle=0]{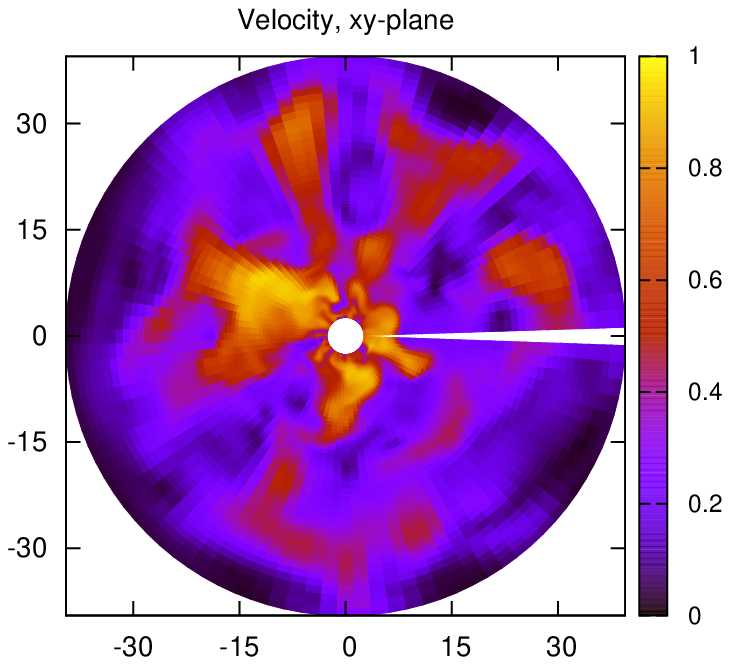} \\
\includegraphics[height=7cm,angle=0]{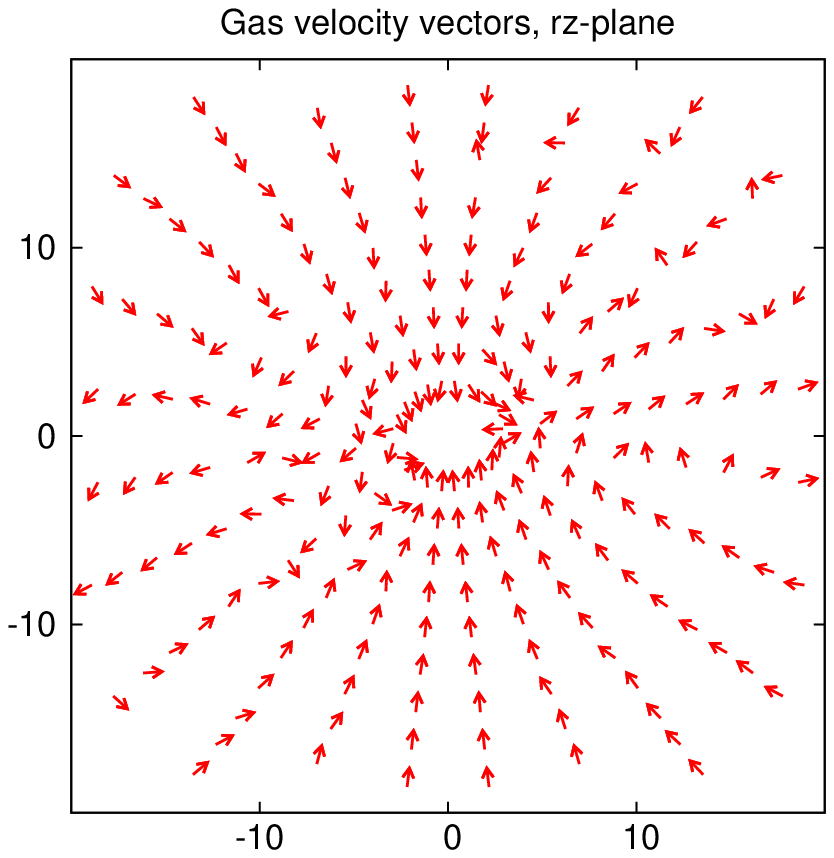} \hspace{.3cm}
\includegraphics[height=7cm,angle=0]{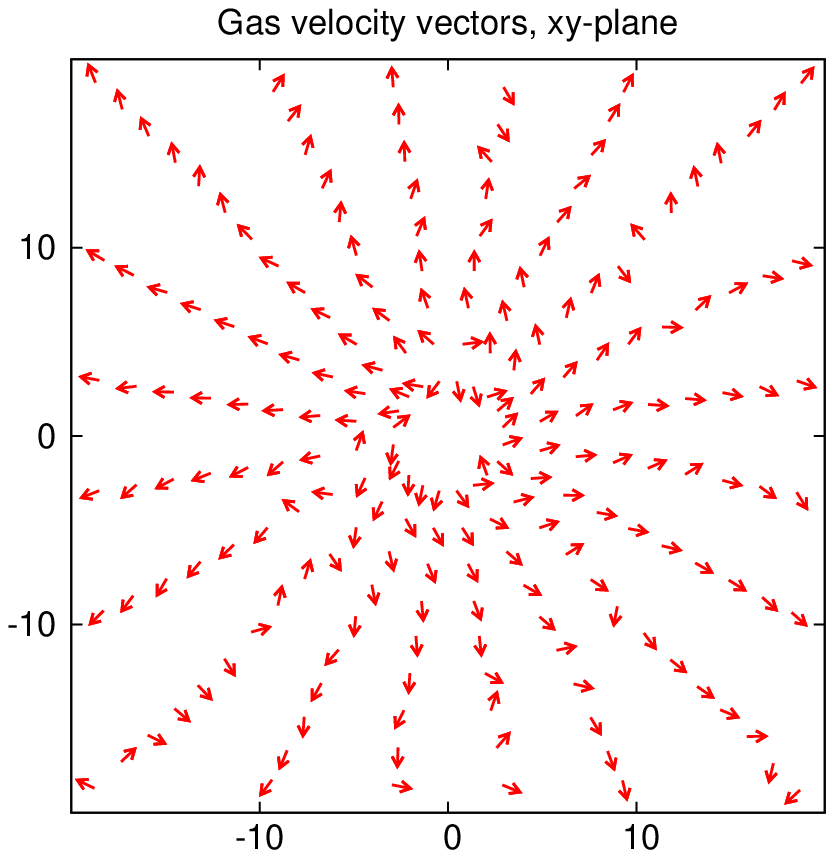}
\end{center}
\par
\vspace{-5mm} 
\caption{As in Fig.~\ref{f-spin0}, in the case of a Kerr 
super-spinar with $a_* = 3$ and at $t = 300 \, M$.}
\label{f-spin3}
\end{figure}

For $a_* = 3$, the accretion process is significantly different
from the cases discussed in the previous subsections.
The difference can be clearly seen in Fig.~\ref{f-spin3}. 
The gas can reach the surface of the super-spinar
from the poles, while the region around the equatorial plane
is characterized by low-density, high-temperature outflows. 
This confirms the results found in~\cite{bhty10} and shows the
behavior of the gas along the third spatial coordinate.
We do not find the formation of collimated and stable outflows,
but a quite chaotic process of ejection of gas. The maximum
temperature of the gas is in the range $400 - 700$~MeV, but most
of the gas is much cooler. The velocity of the gas in the outflows
is found to be close to 1.

\section{Discussion \label{s-disc}}

On the basis of our simulations, we can see that the accretion
process is more efficient onto objects with small spin parameter.
As $|a_*|$ increases, it gets more and more difficult to make all
the accreting gas be swallowed by the compact object. Eventually
the repulsive gravitational force becomes strong enough to produce 
powerful equatorial outflows. In these 3D simulations we do not 
find a perfect axial symmetry, but find that the formation
of outflows is a quite chaotic phenomenon. These are the main 
features of accretion in Kerr space-time.

\begin{figure}
\par
\begin{center}
\includegraphics[height=5.7cm,angle=0]{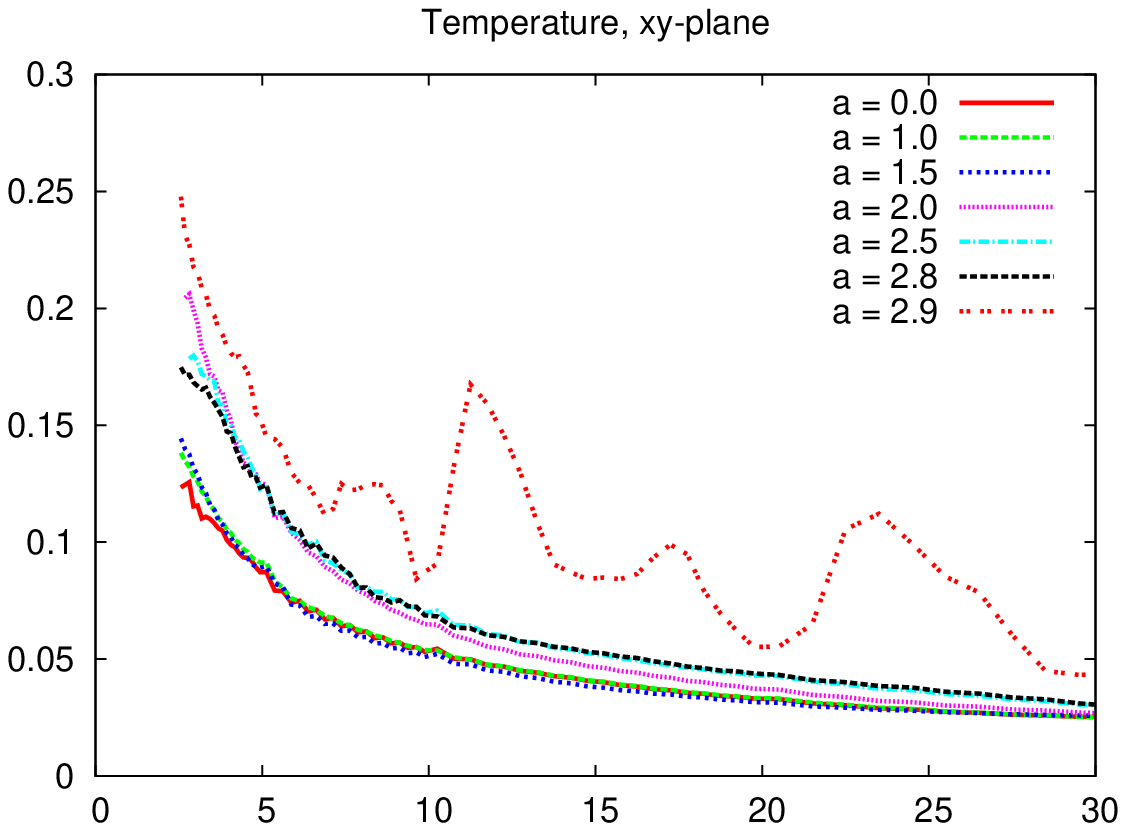} \hspace{.3cm}
\includegraphics[height=5.7cm,angle=0]{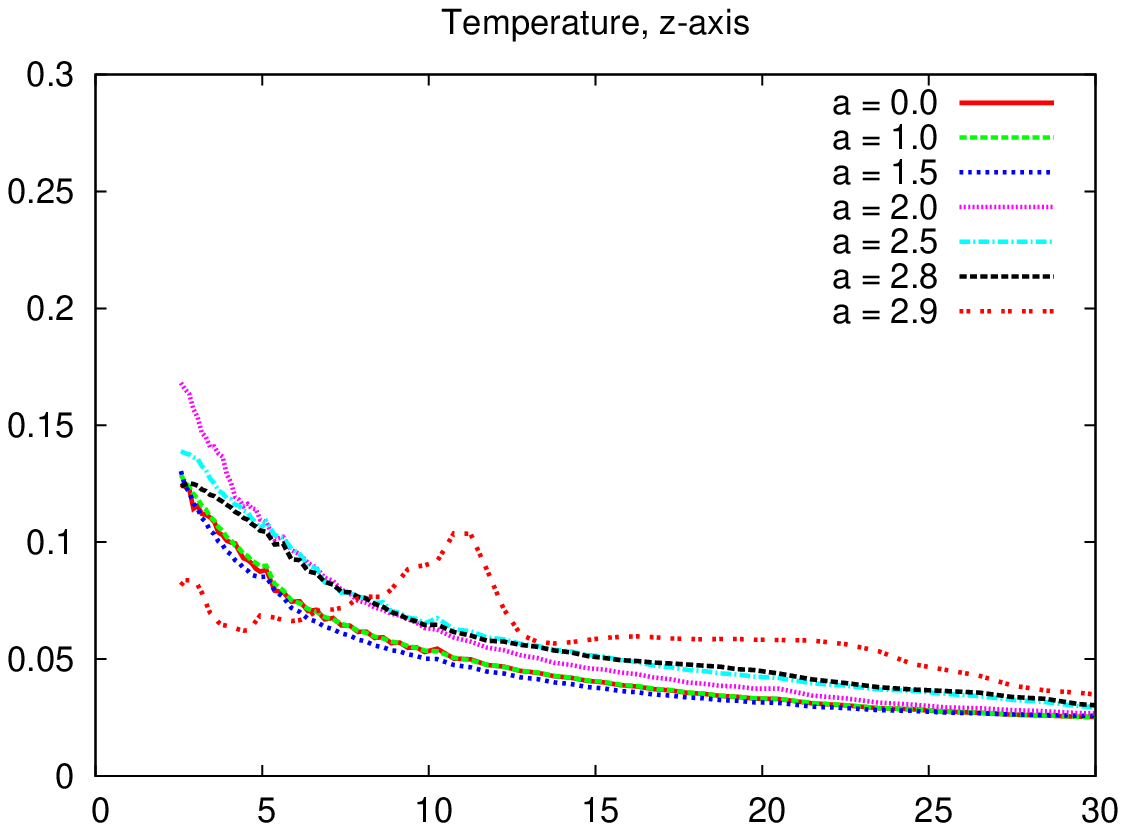} \\
\includegraphics[height=5.7cm,angle=0]{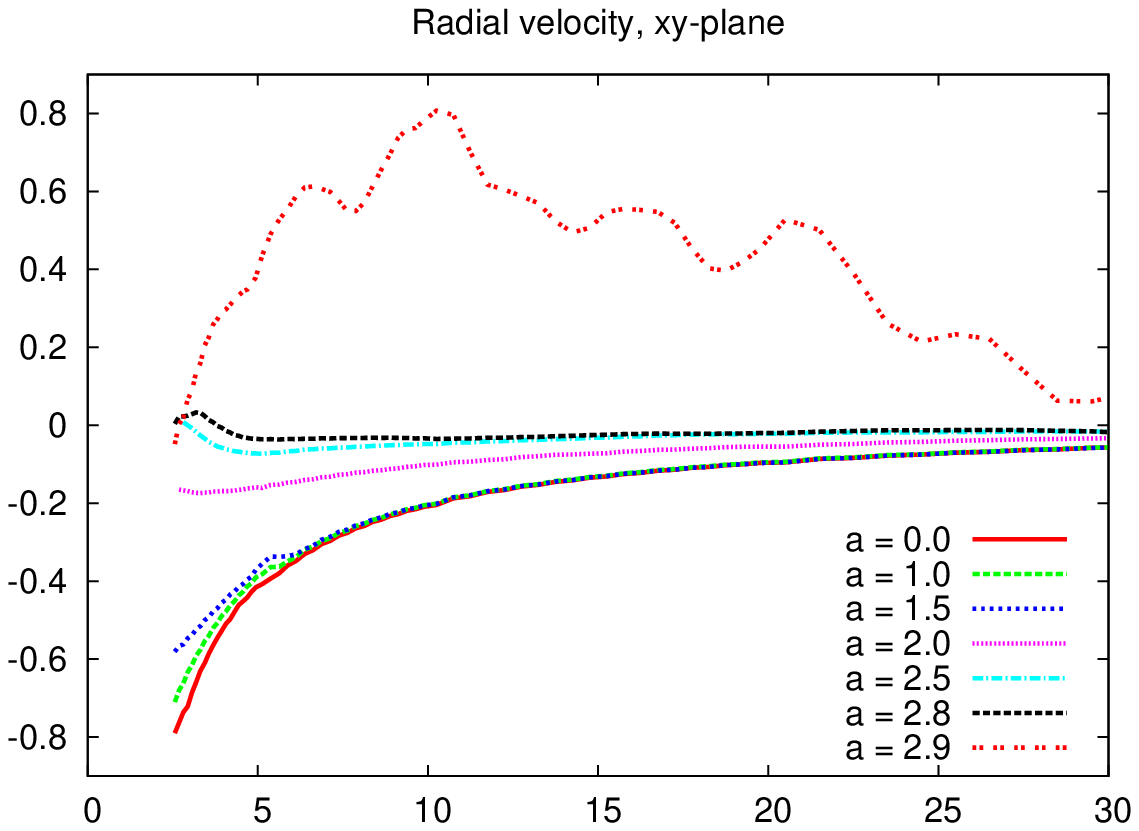} \hspace{.3cm}
\includegraphics[height=5.7cm,angle=0]{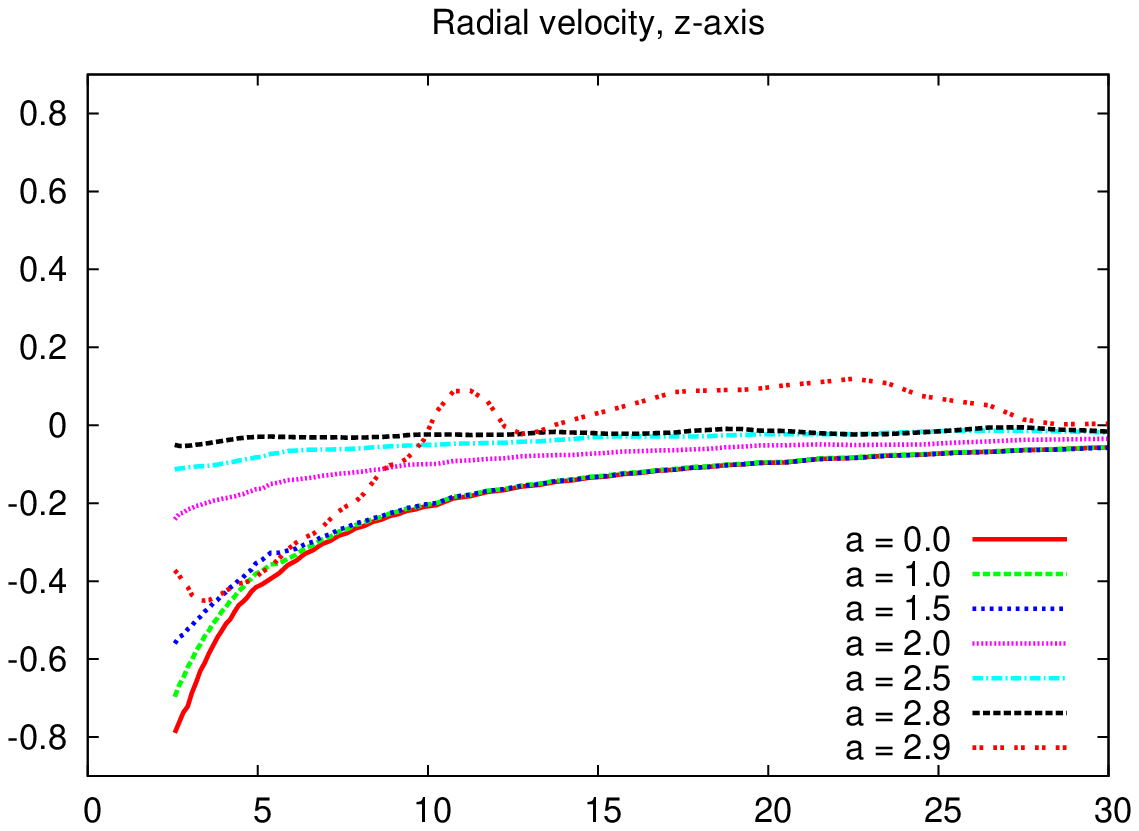}
\end{center}
\par
\vspace{-5mm} 
\caption{Temperature and radial velocity (as seen by a local
observer) as a function of the radial coordinate on the 
equatorial plane ($xy$-plane) and along the $z$-axis of the 
accreting gas in Kerr space-time at $t = 500 \, M$, for 
different values of $a_*$. $r_{in} = 2.5 \, M$, temperature 
in GeV, radial coordinate in units $M = 1$. For $a_* = 0$, 1,
and 1.5, we find a black hole-like accretion; for $a_* = 2$,
an intermediate accretion; for $a_* = 2.9$, the accretion is
of super-spinar type. For $a_* = 2.5$ and 2.8, the accretion 
is essentially of the second kind, but there is some very
weak ejection of matter near the equatorial plane.}
\label{f-prof}
\end{figure}

The transition between the three qualitatively different states
(black hole-like state, intermediate state, super-spinar-like state) 
is rapid, but not instantaneous. To see this, in Fig.~\ref{f-prof} we 
show the temperature and the velocity profile on the equatorial plane and 
along the axis of symmetry for $r_{in} = 2.5 \, M$ and at $t = 500 \, M$.
The cases $a_* = 0$, 1, and 1.5 are quite similar. When $a_* = 2$,
the attractive gravitational force around the massive object is 
not strong and the accretion process is less efficient: 
the velocity of the gas decreases significantly,
while the temperature increases. The flow around the massive
object becomes subsonic. For $a_* \sim 2.5$, the region with
repulsive gravitational force appears in the computational domain.
We then see weak outflows on the equatorial plane. The outflow
is so weak that it cannot go far from the massive object, but is
instead pushed back by the accreting gas. As the spin parameter
increases, the energy of the outflows increases as well. For 
$a_* = 2.9$, the outflows can reach the outer boundary of the 
computational domain at $r_{out} = 40 \, M$. 
We have checked that, for a smaller/larger $r_{in}$, 
qualitatively the same behavior is found 
with a lower/higher value of the spin parameter. For
example, setting $r_{in} = 3.0 \, M$, weak equatorial outflows 
start forming for $a_* \sim 3.5$ and they becomes powerful when
$a_* = 3.8$. The two key elements of the accretion process are
thus the spin parameter and the radius of the massive object. 
The latter is replaced by the radius of the inner boundary in
our simulations. A minor role is played by the initial and 
boundary conditions, like the temperature and the velocity of 
the gas, i.e. by non-gravitational physics.

\begin{figure}
\par
\begin{center}
\includegraphics[height=5.7cm,angle=0]{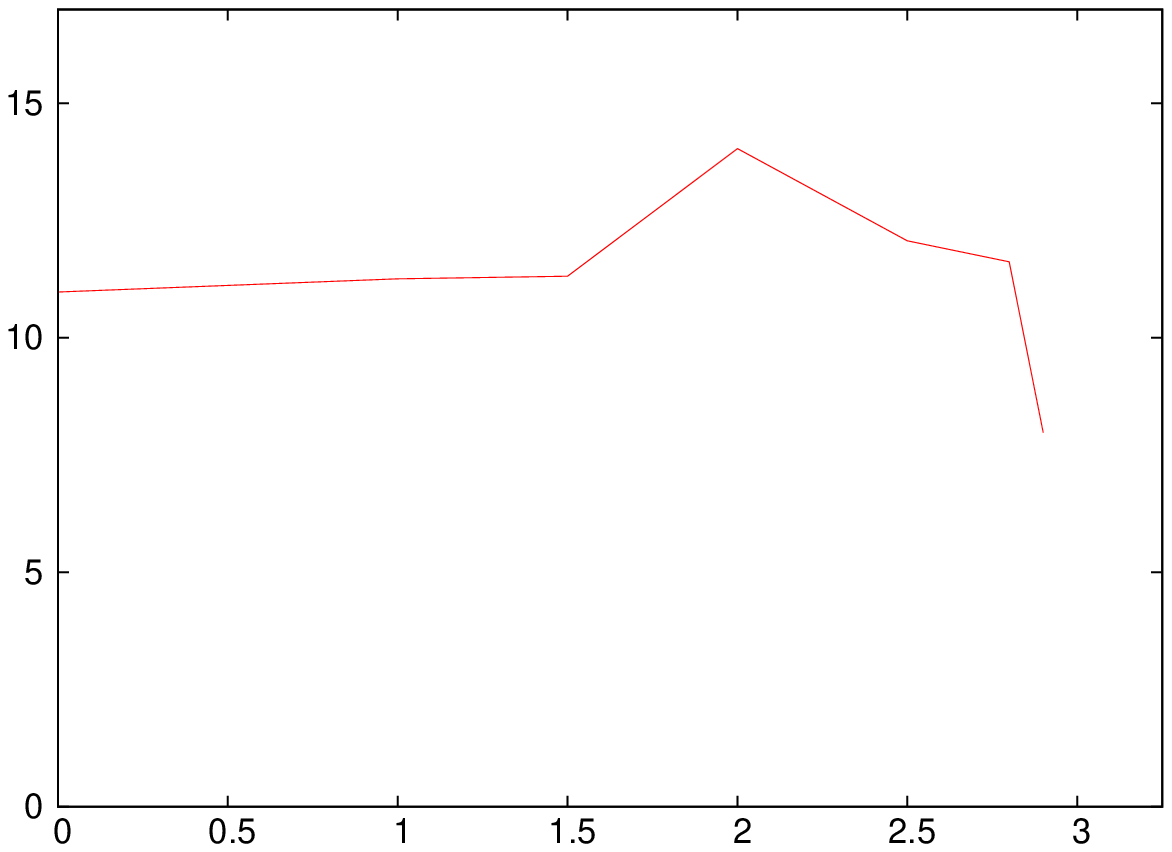} \hspace{.3cm}
\includegraphics[height=5.7cm,angle=0]{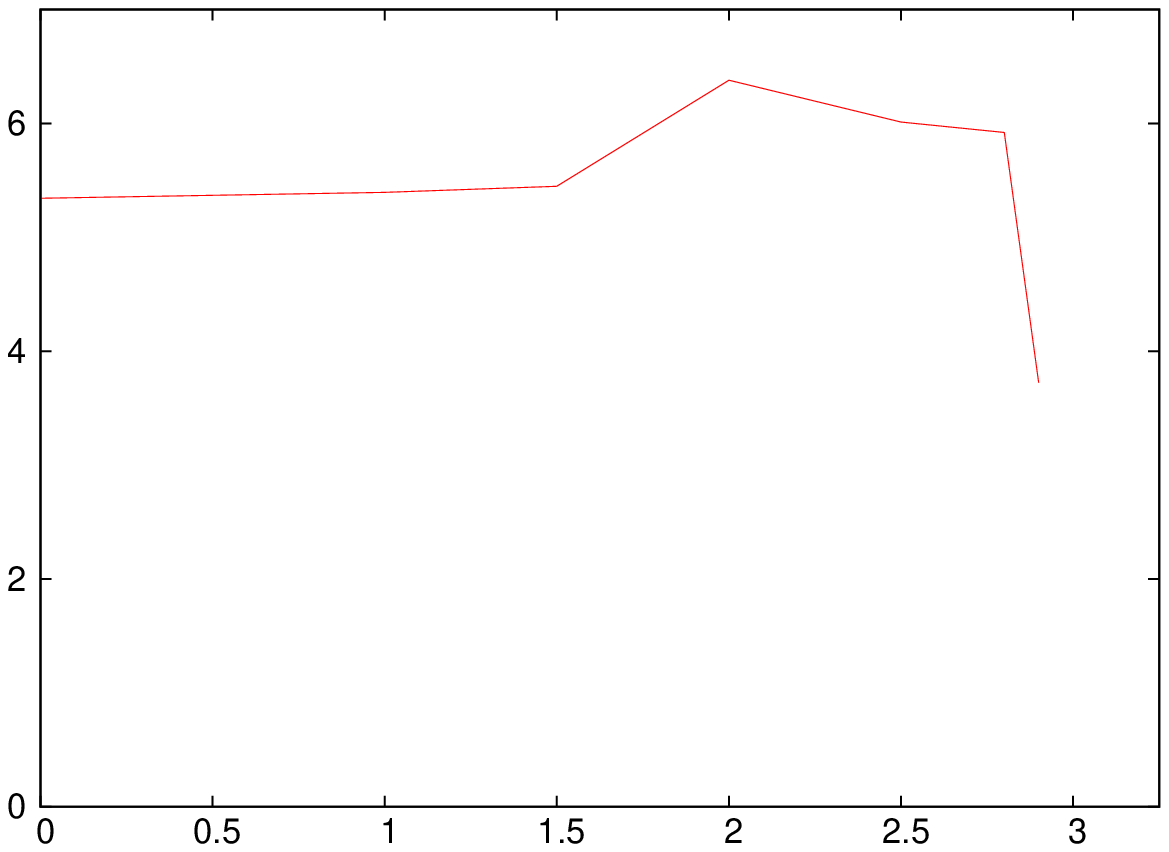}
\end{center}
\par
\vspace{-5mm} 
\caption{Mean density of the accreting gas (in arbitrary units) 
around the massive object at $t = 500 \, M$ as a function 
of the spin parameter $a_*$. Left panel: space region
$r_{in} < r < 5 \, M$. Right panel: space region 
$r_{in} < r < 10 \, M$.}
\label{f-mass}
\end{figure}

The mass of the accreting gas around the massive object, $M_{gas}$, 
or correspondingly the mean density $\rho_{gas}$, could be used as order
parameter of the system. In the black hole state of accretion, 
these two quantities slightly increase as $a_*$ increases. In the
intermediate state, the mass and the density around the massive
object reach a maximum, while, as soon as outflows can be produced, they
drop to lower values. In Fig.~\ref{f-mass}, we show, as a function 
of the spin parameter $a_*$, the mean density of the accreting gas in 
the region $2.5 \, M < r < 5.0 \, M$ (left panel) and 
$2.5 \, M < r < 10.0 \, M$ (right panel) and at the time $t = 500 \, M$.
This fact can, however, unlikely be used in observations, because 
the mass and the density of the gas around the compact object are
mainly determined by the accretion rate at larger distances.
Here we simply note that, for the same initial and boundary 
conditions, the spin of the massive object determines the amount
of gas around the center as shown in Fig.~\ref{f-mass}.

In our super-spinar model, the surface of the massive object
absorbs all the accreting matter. Namely, we have assumed
a transmission coefficient $T = 1$ and, consequently, a reflection 
coefficient $R = 1 - T = 0$. In this case, the surface of the
super-spinar would be similar to an event horizon, in the sense
that it can swallow all the matter with no difficulties. The
properties of the surface of the super-spinar should depend on
how the ordinary matter of the gas, made of protons and electrons, 
interacts with the exotic structure of the super-spinar. 
The opposite case of a perfectly absorbing surface is a perfectly
reflecting surface with $T = 0$ and $R = 1$ (rigid wall). To see the 
behavior of the accretion process with radically different 
properties of the surface of the super-spinar, we run the code
imposing reflective boundary conditions at $r_{in}$. For black holes
and super-spinars with low spin parameter, we found a quite obvious
result: 
a dense cloud is formed around the massive object,
which quickly explodes. For super-spinars with moderate
and high spin parameter, the system is much more stable, thanks to the
non-spherical symmetry of the gravitational field. Now there is not
the formation of the unstable cloud, since the gas around the 
massive object is efficiently expelled in the outflows: the gas
reaches the surface of the super-spinar from the poles, there is
no accretion at all, because of the perfectly reflecting surface, 
and is ejected around the equatorial plane. We argue that the final
result is the formation of a convective zone, where the gas
continuously approaches and leaves the super-spinar. In a more
general case, one could expect that either $T$ and $R$ are non-zero,
smaller than 1, and depend on the energy of the incident particle,
i.e. $T = T(\omega)$ and $R = R(\omega)$. In any case, outflows
on the equatorial plane for high spin parameters seem to be a
quite robust prediction of the accretion process onto super-spinars,
regardless of the details of their surface. 
Indeed the assumption of a perfectly absorbing surface is
perhaps the most conservative possibility; that is, the case in which it
is more difficult to produce equatorial outflows.

\begin{figure}
\par
\begin{center}
\includegraphics[height=5.7cm,angle=0]{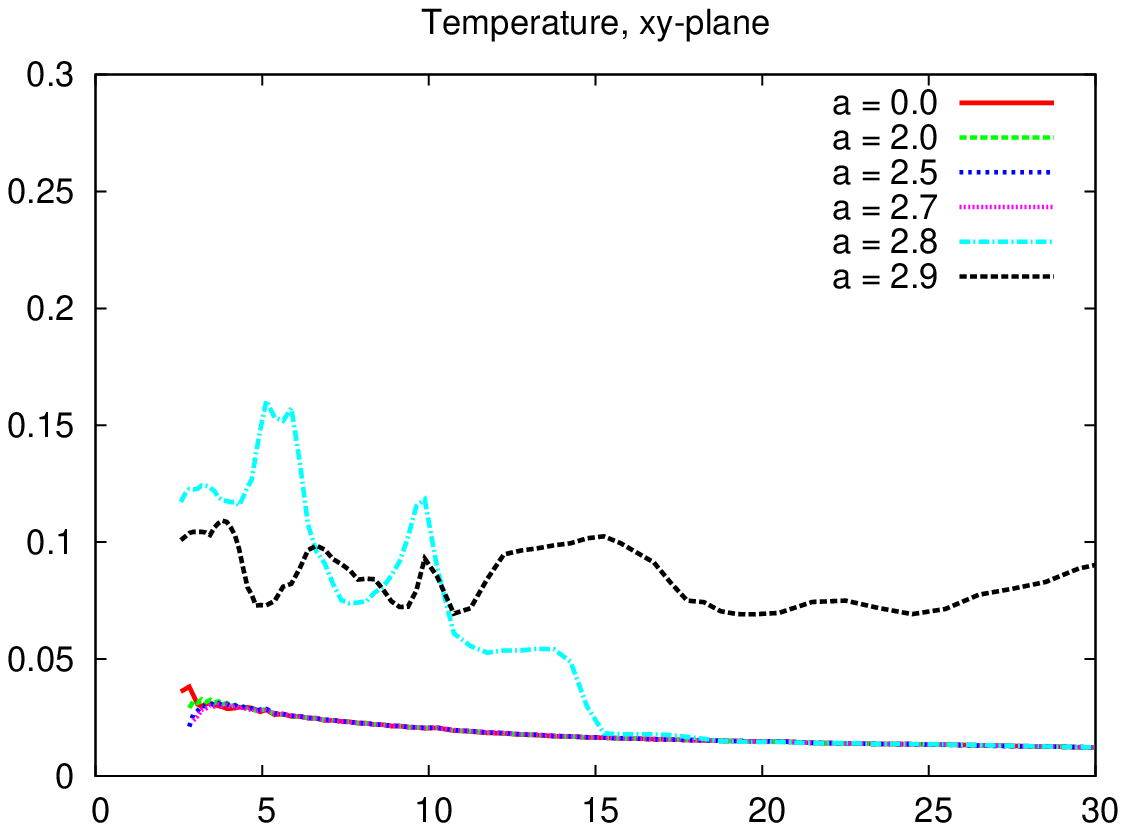} \hspace{.3cm}
\includegraphics[height=5.7cm,angle=0]{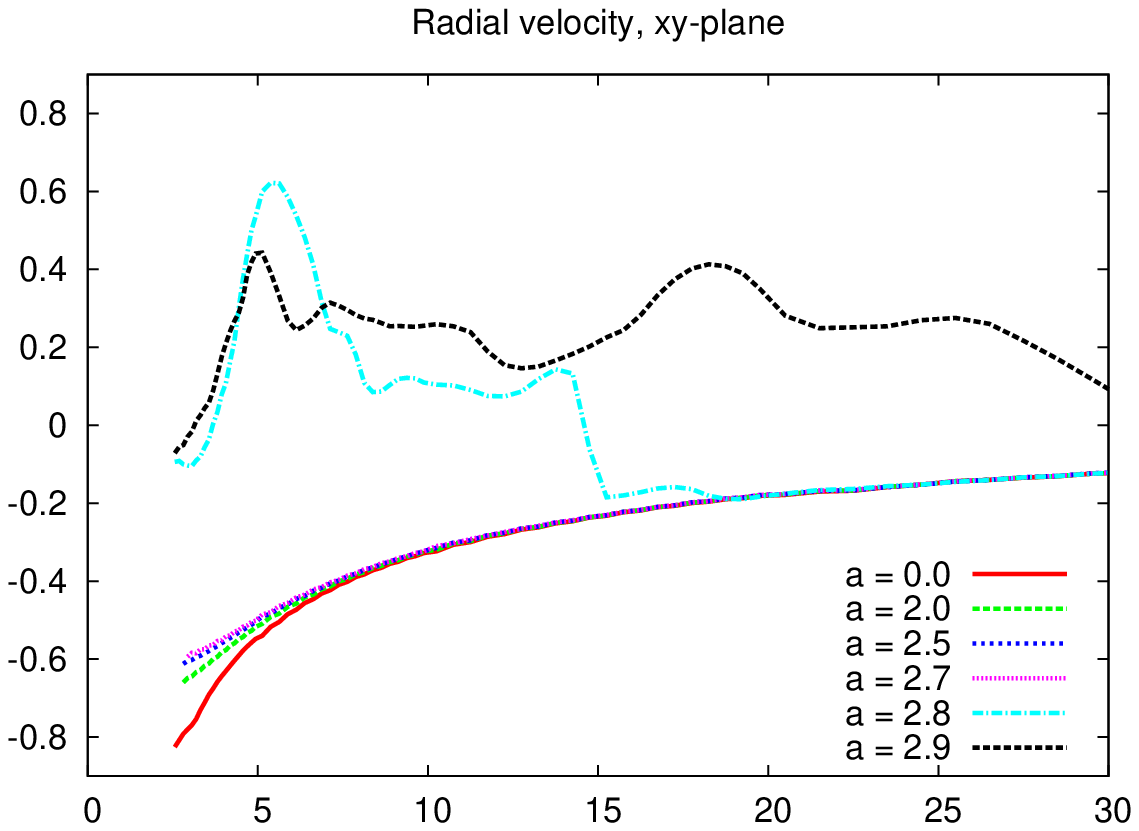}
\end{center}
\par
\vspace{-5mm} 
\caption{Temperature and radial velocity (as seen by a local
observer) as a function of the radial coordinate on the 
equatorial plane ($xy$-plane) of the accreting gas in Kerr 
space-time at $t = 500 \, M$, for different values of $a_*$,
in the case $\Gamma = 4/3$.}
\label{f-profrel}
\end{figure}

Lastly, we have briefly investigated the behavior of the accretion
process for a gas with a different equation of state. So far, in our 
study we have always assumed that the gas consists
non-relativistic particles; that is, $\Gamma = 5/3$. On the other 
hand, around an ordinary black hole, one should expect non-relativistic
ions and relativistic electrons. This fact could be easily 
implemented in our code by taking $\Gamma = 13/9$. Since 
here we want only to understand how our results are affected by 
the gas equation of state, we consider the more radical case of a 
gas whose pressure is dominated by relativistic particles,
i.e., $\Gamma = 4/3$. 
The accretion process turns out to be qualitatively 
the same. Quantitatively, there are some differences, but that should 
not be a surprise, because even the standard Bondi accretion onto a
Schwarzschild black hole depends on the matter equation of state.
The results of our simulations for $\Gamma = 4/3$ are summarized
in Fig.~\ref{f-profrel}, where we show the temperature and radial
velocity profile on the equatorial plane for a few different values
of the spin parameter $a_*$. Fig.~\ref{f-profrel} should be
compared with Fig.~\ref{f-prof}. For $r_{in} = 2.5 \, M$, the
accretion process is essentially the same for any value $|a_*| \le 2.7$.
For $a_* = 2.8$, there are equatorial outflows, but the gas cannot
go far from the massive objects. For $a_* = 2.9$, the outflows
are apparently less energetic than the ones when $\Gamma = 5/3$.

\section{Conclusions \label{s-conc}}

A Kerr black hole must satisfy the relation $|a_*| \le 1$,
where $a_* = J^2/M$ is the dimensionless spin parameter.
So, the possible discovery of a massive and compact object
with $|a_*| > 1$ would imply that the final product of the
gravitational collapse is not a Kerr black hole, or at least
that the Kerr solution is not the unique option. To test 
the bound $|a_*| \le 1$ in astrophysical black hole candidates 
with electromagnetic radiation, we have to study the accretion 
process onto compact object either with $|a_*| \le 1$ (black 
holes) and with $|a_*| > 1$ (super-spinars).

In this paper, we have presented the results of our 3-dimensional
general relativistic hydrodynamic simulations of adiabatic and
spherically symmetric accretion in Kerr space-time. Kerr 
super-spinars have been modeled as a compact object with radius 
$r \approx 2.5 \, M$ and a surface made of exotic stuff capable 
of absorbing the matter hitting it: from the one hand, this choice
is motivated to prevent the instability of the massive object 
and, from the other hand, it is inspired by a couple of scenarios 
proposed in the literature.

Our simulations in 3 dimensions suggest three main regimes of 
accretion: black hole-like state, intermediate state, and 
super-spinar-like state. The key elements determining the kind 
of accretion are the spin parameter $a_*$ and the radius of the 
massive object -- here the radius of the inner boundary $r_{in}$. 
However, for a smaller/larger radius, we can rescale 
(decrease/increase) the spin parameter and obtain a similar 
accretion process. For black holes and super-spinars with small 
$|a_*|$, we find the usual picture of accretion onto black holes,
with the thermodynamical variables quite independent of the 
actual value of $a_*$ (black hole-like state). For super-spinars 
with moderate value of $|a_*|$, the flow around the massive object 
becomes subsonic and the accretion process is more difficult, 
making the density and the temperature of the gas increase
(intermediate state). For high values of $|a_*|$, the accretion 
process is quite different: the super-spinar accrete from the 
poles, while the equatorial plane is dominated by powerful 
outflows (super-spinar-like state).

These results confirm the findings in our previous works in 
2.5 dimensions~\cite{bhty10}. Now we can also see the details
of the structure 
of the outflows on the equatorial plane. Although our simulations
do not run for enough time to reach a quasi-steady-state equilibrium
configuration in the case of high spin parameter, we argue that 
the accretion process onto such super-spinars is characterized by the
random ejections of gas in essentially all the directions, without the 
formation of stable structures.


\begin{acknowledgments}
We would like to thank Tomohiro Harada and Rohta Takahashi
for collaboration on the early stage of this work. We are
also grateful to Enrico Barausse for useful discussions
about the issue of the stability of super-spinars.
The work of C.B. was partly supported by the JSPS 
Grant-in-Aid for Young Scientists (B) No. 22740147.
This work was supported by World Premier International 
Research Center Initiative (WPI Initiative), MEXT, Japan.
\end{acknowledgments}


\end{document}